\documentclass[twocolumn,showpacs,preprintnumbers,floatfix,prl]{revtex4}
\usepackage[usenames]{color}
\usepackage{graphicx}
\usepackage{amsmath}

\begin{document}

\title{Experimental study of the compaction dynamics for 2D anisotropic granular materials}

\author{G. Lumay and N. Vandewalle}

\affiliation{GRASP, Institut de Physique B5, Universit\'e de Li\`ege, \\ B-4000 Li\`ege, Belgium.}

\begin{abstract}

We present an experimental study of the compaction dynamics for two-dimensional anisotropic granular systems. The compaction dynamics of rods is measured at three different scales : (i) the macroscopic scale through the packing fraction $\rho$, (ii) the mesoscopic scale through both fractions of aligned grains $\phi_{a}$ and ideally ordered grains $\phi_{io}$, and (iii) the microscopic scale through both rotational and translational grain mobilities $\mu_{r,t}$. At the macroscopic scale, we have observed two stages during the compaction process suggesting different characteristic time scales for grain relaxation. At the mesoscopic scale, we have observed the formation and the growth of domains made of aligned grains during the first stage of compaction. At the late stage, these domains of aligned grains are sheared to form ideally ordered domains. From a microscopic point of view, measurements reveal that the beginning of the compaction process is essentially related to translational motion of the grains. The grain rotations drive mainly the process during the late stages of compaction.

\end{abstract}

\maketitle


\section{Introduction}

Most of the industrial products are processed, transported and stocked in a granular state. The packing fraction of those granular materials becomes therefore a relevant parameter for a broad range of applications. The best way to reduce the costs for the manipulation of such granular materials is to increase the packing fraction $\rho$. This can be achieved by tapping or vibrating the vessel containing the grains. 

Granular matter has been the subject of numerous studies since the last decade \cite{deGenne,granularSolidLiquidsGases,Duran,Kudrolli,Francois}. From a scientific point of view, granular system are interesting system because they exhibit disorder, complex behaviors and strong analogies with glasses \cite{Glasses} and colloids \cite{vanMegen}. 

Various experimental studies \cite{densFluc,densityRelax,kww1,richard,LumayVandewalle2} have underlined the fact that the dynamics of compaction is a complex problem. The compaction is actually characterized by an extremely slow dynamics. Different laws have been proposed for the volume fraction $\rho$ of a granular assembly as a function of the number $n$ of taps. Since both initial $\rho_0$ and final $\rho_\infty$ packing fractions depend on the experimental conditions, a normalized parameter is defined. One has

\begin{equation}
\tilde{\rho}=\frac{\rho-\rho_{0}}{\rho_{\infty}-\rho_{0}},
\end{equation} with $0 \leq \tilde{\rho} \leq 1$. This parameter has the great advantage to be reproducible.

It has been proposed by Knight \emph{et al.} \cite{densityRelax}, that the packing fraction obeys an inverse logarithmic law 
\begin{equation}
\label{eq:invLog}
\tilde{\rho}(n) = 1 - \frac{1}{1 + B \ln(1+\frac{n}{\tau})},
\end{equation} where $B$ and $\tau$ are dimensionless parameters. This inverse logarithmic law was also obtained in numerical models like the Tetris one \cite{Tetris,Tetris2}. The law (\ref{eq:invLog}) could also be derived from theoretical arguments \cite{slowRelax}.

Some theoretical models \cite{boutreux,arenzon1,arenzon2} are based on a relationship between the mobility $\mu$ of the grains and the global packing fraction $\rho$ of the packing. The mobility is a local property of the grains and corresponds to the grain ability to move inside the packing. By considering that the variation of the packing fraction induced by a tap is proportional to the grain mobility, one can write the simple equation

\begin{equation}
\label{eq:rhoEvol}
{\partial \tilde{\rho} \over \partial t} = k \; \mu,
\end{equation} where $k$ is a constant. By considering a caging effect of the grains, some authors \cite{boutreux} proposed the Vogel-Fulcher law for the decrease of the mobility $\mu$ with the packing fraction $\rho$. One has

\begin{equation}
\label{eq:vogelFulcher}
\mu = \mu_0 \exp \left( - \frac{c}{1 - \tilde{\rho}} \right).
\end{equation} This relation could be combined with Eq. (\ref{eq:rhoEvol}) in order to obtain an inverse logarithmic behavior for $\tilde{\rho}(n)$.

The slow dynamics of granular compaction has also been described by a cluster model \cite{clusterModel}. A cluster is a group of grains ideally packed. The granular material is then considered as a system of various clusters competing in a random environment. Vibrations cause the slow growth of the cluster size. This growth leads to a logarithmic law (\ref{eq:invLog}) for the evolution of the packing fraction. The relevant parameter for measuring any grain ordering is the fraction $\phi$ of grains ideally packed. This parameter could also be normalized. One has

\begin{equation}
\tilde{\phi}=\frac{\phi-\phi_{0}}{\phi_{\infty}-\phi_{0}},
\end{equation} with $0 \leq \tilde{\phi} \leq 1$. 

More recently, Philippe and Bideau \cite{kww1} found that the compaction dynamics is better fitted by a stretched exponential law, like
\begin{equation}
\label{eq:KWW}
\tilde{\rho} (n) = 1 - \exp\left[  - \left(\frac{n}{\tau}\right)^{\beta} \right].
\end{equation} This exponential law presents the great advantage to fit a saturation of the packing fraction which is sometimes accessible in experiments (for larges $n$ values). Both parameters $\tau$ and $\beta$ correspond respectively to a characteristic tap number and to a stretching exponent. One should notice that in the experiments of Philippe and Bideau, the steady state corresponds to a dynamical balance between convection and compaction.

In a previous experimental work \cite{LumayVandewalle2}, we have performed a multiscale study of the compaction dynamics for a 2D pile of spherical particles. We have shown that granular compaction dynamics could be viewed as a slow process of crystallization driven by the diffusion of defects in the packing. The evolution of the normalized fraction of ideally ordered grains $\tilde{\phi}$ is well described by the Avrami model. One has

\begin{equation}
\label{eq:AvramiLaw}
\tilde{\phi} = 1-\exp \left[ - \left( \frac{n}{\tau}\right) ^ {\gamma} \right].
\end{equation} The Avrami exponent $\gamma$ depends on the nature of the growth and $\tau$ is a characteristic time. For the compaction of spherical particles, we have obtained an exponent $\gamma = 0.5$. A new law for the compaction dynamics can be derived from this crystallization model. One has

\begin{equation}
\label{eq:AvramiRhoLaw}
\tilde{\rho} (n) = \sqrt{ 1 - \exp \left(  - \sqrt{\frac{n}{\tau}} \right)},
\end{equation} \textit{i.e.} a law similar to the Philippe and Bideau's law. The combination of both Eq (\ref{eq:rhoEvol}) and (\ref{eq:AvramiRhoLaw}) gives a law for the mobility. One has

\begin{equation}
\label{eq:AvramiMuLaw}
\mu = -\frac{1}{k}\frac{1-\tilde{\rho}^2}{\tilde{\rho} \; \ln{(1-\tilde{\rho}^2)}}.
\end{equation} 

In a couple of previous papers \cite{PG2005Lum,Tshape}, we proposed an empirical law to fit the decrease of the mobility $\mu$ with the normalized packing fraction $\tilde{\rho}$, instead of Eq. (\ref{eq:AvramiMuLaw}). One has

\begin{equation}
\label{eq:EmpiricalMuLaw}
\mu = \mu_{0} (1-\tilde{\rho}) e^{-a \tilde{\rho}}
\end{equation} where $\mu_{0}$ is a constant initial mobility and $a$ is the decrease rate of the mobility with the normalized packing fraction $\tilde{\rho}$. This exponential law presents a practical interest. Indeed, the value of the parameters $a$ and $\mu_{0}$ could be easily fitted and discussed. 

The Table \ref{tab:spheric} summarizes the different laws proposed for describing the compaction of isotropic granular materials. 

\begin{table*}
\caption{Summary of the laws proposed by different authors in order to describe the compaction dynamics of isotropic granular materials. The symbol $*$ means that a numerical resolution of the equation (\ref{eq:rhoEvol}) is necessary. The cross $\times$ means that the crystallization is not observed.  }
\begin{ruledtabular}
\begin{tabular}{llllll}
\hline Granular type & packing fraction & crystallization & grain mobility & comment & ref\\ 
\hline 
3D pile of spheres & 
$\tilde{\rho} = 1 - \frac{1}{1 + B \ln(1+\frac{n}{\tau})}$ & 
$\times$ & 
$\mu = \mu_0 \exp \left( - \frac{c}{1 - \rho} \right) $& 
semi-empirical law & 
\cite{densFluc,densityRelax}\\ 
\hline 
3D pile of spheres & 
$\tilde{\rho} = 1 - \exp\left[  -\left(\frac{n}{\tau}\right)^{\beta} \right] $& 
$\times$ & 
$*$& 
semi-empirical law & 
\cite{kww1}\\ 
\hline 
2D pile of disks & 
$\tilde{\rho} = \sqrt{ 1 - \exp \left(  - \sqrt{\frac{n}{\tau}} \right)}$ &
$\tilde{\phi} = 1-\exp \left[ - \left( \frac{n}{\tau}\right) ^ {1/2} \right]$& 
$\mu = -\frac{1}{k}\frac{1-\tilde{\rho}^2}{\tilde{\rho} \; \ln{(1-\tilde{\rho}^2)}}$& 
theoretical law & 
\cite{LumayVandewalle2}\\ 
\hline 2D pile of disks & 
$*$& 
$\times$ &
$\mu = \mu_{0} (1-\tilde{\rho}) e^{-a_{t} \tilde{\rho}}$ & 
empirical law & 
\cite{PG2005Lum,Tshape}\\
\hline 
\end{tabular} 
\end{ruledtabular}
\label{tab:spheric}
\end{table*}


Granular materials are rarely composed of perfectly spherical particles. Recently, some studies on 3D packing of anisotropic grains have been performed \cite{Kudrolli2, LumayVandewalle1, convectAnisotropic, rods, thin-rod, spherocylinders}. The major difference between sphere and cylinder packings is the tendency of cylinders to align themselves along their symmetry axis. Cylinders have also the tendency to align themselves along the container walls. Villaruel \emph{et al.} \cite{rods} studied the compaction of cylindrical particles with an aspect ratio $\alpha = 3.9$ in a thin tube. In their three-dimensional experiment, the ratio between the particle length and the container diameter is $\ell/D = 2.7$. During the compaction, they observe three stages. Underlining that the previous laws (1-10) capture only a part of the physical mechanisms behind compaction. The first stage is a rapid vertical collapse of the pile. The second stage is a slow ordering of the particles along the sidewalls. The final stage is a steady state. In a previous work \cite{LumayVandewalle1}, we have shown the importance of the grain aspect ratio on the compaction properties for three-dimensional packing. Stokely \emph{et al.} \cite{prolatePacking} investigated a 2D packing of extremely prolate granular material (aspect ratio $\alpha > 10$). They observed a strong orientational correlation for particles separated by less than two particle lengths. Furthermore, a general preference for horizontal alignment was observed.

In the present paper, we investigate the compaction of a 2D pile of cylindrical particles. In the first part of the paper, we give some details about the tap characteristics. Then, the compaction dynamics is analyzed at three different scales : (i) the macroscopic scale through the normalized packing fraction $\tilde{\rho}$, (ii) the mesoscopic scale through both fractions of grains aligned $\phi_{a}$ and ideally ordered grains $\phi_{io}$, and (iii) the microscopic scale through the grain rotational and translational mobilities $\mu_{r,t}$.


\section{Experimental set-up}

\begin{figure}[h]
\begin{center}
\includegraphics[scale=0.3]{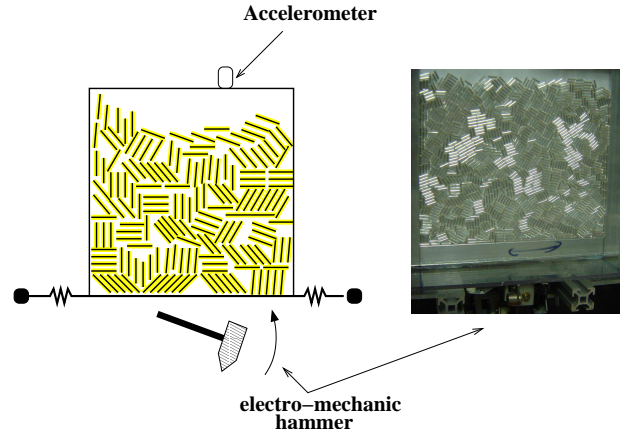} 
\end{center}
\caption{A sketch (left) and a picture (right) of our experimental setup. Approximately 1200 cylinders are placed between two parallel plates. An electro-mechanic hammer is placed below the pile and is tuned by a micro-controller. An accelerometer is placed on the vessel to measure the acceleration experienced by the whole system.}
\label{im:setup}
\end{figure}

Cylindrical particles are placed between two vertical parallel plates. The number of particles is typically $N \approx 1200$. The cylinder diameter is $d = 2.16 \pm 0.05$ mm and the cylinder length is $\ell = 6.73 \pm 0.16$ mm. Therefore, the aspect ratio of the cylinder is $\alpha = 3.1$. In order to avoid the possible overlapping of the grains, the distance between both vertical plates is slightly greater than the particle diameter. The width of the pile is 150 mm ($\approx  22 \ell$) and the mean height is initially 140 mm ($\approx  20 \ell$). Piles are formed by (1) randomly distributing particles on one glass plate, (2) fixing the second plate, and (3) slowly tilting the system to the vertical. A sketch and a picture of the setup are given in Figure \ref{im:setup}. 

\begin{figure}[h]
\begin{center}
\includegraphics[scale=0.4]{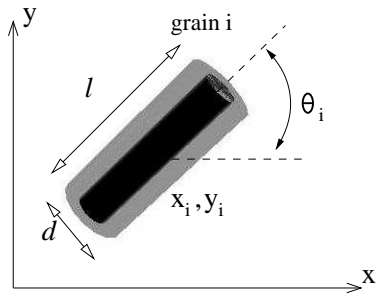} 
\end{center}
\caption{A sketch of a grain labelled $i$ characterized by its position $(x_i,y_i)$ and its orientation $\theta_i$. Each grain is a metallic cylinder enclosed in a glass cylinder.}
\label{im:grain}
\end{figure}

The grains we used are formed by a metallic rod enclosed in a glass cylinder (see Figure \ref{im:grain}). The pile is rear illuminated by an homogeneous light source. In such a way, the metallic core of the grains appears black and their outline appears bright. With this configuration, parallel grains can be resolved. A camera records pictures of the pile during the experiment. Therefore, we can measure the position ($x_i,y_i$) and the orientation ($\theta_i$) of each grain $i$ by image treatment. The study of the grains motion during a tap is performed by an ultrafast camera. The compaction dynamics is analysed by a high resolution camera ($1024 \time 1024$ pixels). 

To produce the successive taps, an electro-mechanic hammer is placed below the container. The hammer is tuned by a micro-controller that can adjust the intensity, the number and the frequency of the taps. The acceleration experienced by the system at each tap is measured by an accelerometer connected to an oscilloscope. Taps are characterized in the next section. Two successive taps are separated by 500 ms. This time is much longer than the relaxation time of the system after each acceleration peak (see below). Since the dynamics of compaction slows down with the number of taps $n$, we recorded images of the system only for $n = 2^i$ with $i = 1,2,...,18$. In order to improve the measurements, some additional images are recorded at the end of the process.

\section{Tap characteristics}

Similarly to our previous works \cite{LumayVandewalle1,LumayVandewalle2}, we use an original system to produce taps. Some details about the tap characteristics and about the response of the pile are given in this subsection. 

\begin{figure}[h]
\begin{center}
\includegraphics[scale=0.5]{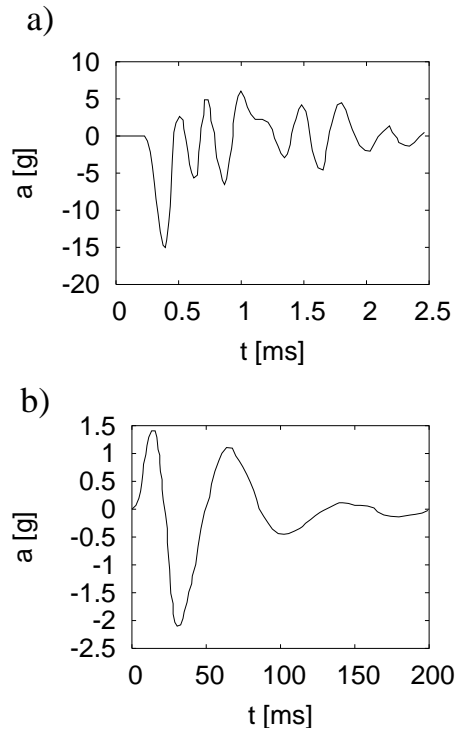} 
\end{center}
\caption{(a) Acceleration experienced by our system during a tap produced by an electro-mechanic hammer. The system undergoes a short peak of acceleration. The width of the main peak is 0.25 ms and the maximum intensity is 15 g. Some damped oscillations are observed during a few milliseconds. (b) Acceleration experienced by the system during a tap produced by an electromagnetic exciter in the Rennes experiments \cite{pierrePhilippe}. }
\label{im:acceleration}
\end{figure}

As we can see in Figure \ref{im:acceleration}a, the system undergoes a short and strong peak of negative acceleration during a tap. The main peak lasts 0.25 ms and the maximum intensity reaches -15g. Some damped oscillations during a few milliseconds are observed. The movement amplitude of the system is very small with respect to others. In both Rennes \cite{richard,kww1} and Chicago \cite{rods,densityRelax,slowRelax} experiments, taps are produced by an electromagnetic exciter and consist of an entire cycle of a sine wave. For high accelerations, the considerable amplitude of the container movement produces some convection in the pile. It should be noted that such a global motion is not present in our experiment. With our set-up, we only observe the compaction phenomenon. 

\begin{figure}[h]
\begin{center}
\includegraphics[scale=0.6]{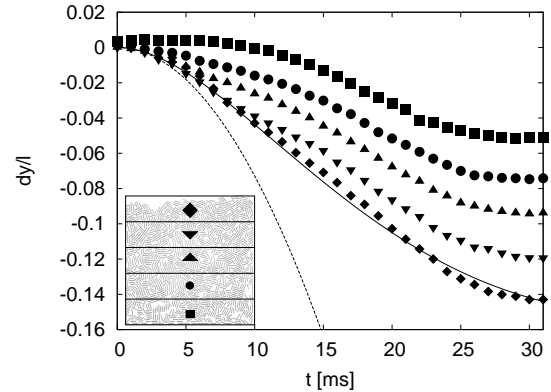} 
\end{center}
\caption{Average vertical displacement $\bar{d y}$ of the grains diveded by the grains lenght $l$ just after the first tap ($n = 1$). The pile is divided in five vertical layers as shown in the inset. The diplacement $\bar{d y}/l$ is shown for each layer. The dashed curve is a free fall. The solid curve is fit by a free fall with a damping mechanism.}
\label{im:deltaYAfterTap1}
\end{figure}

In order to characterize precisely the grain motions induced by a tap, we have recorded the pile with a fast video recorder (1000 images/s) just after the first tap of a series. The position of each grain ($x_{i}$,$y_{i}$) has been measured on each picture. The index $i$ is the number of the grain. The pile has been divide in five horizontal layers (see Figure \ref{im:deltaYAfterTap1} inset). The mean vertical displacement $\bar{dy}$ of the grains just after the tap has been compute in these layers (i.e. for different heights in the pile). For the layer L, one has 
\begin{equation}
\bar{dy}_{L}(t) = \frac{1}{N_{L}}\sum_{i \in L} y_{i}(t)-y_{i}(0), 
\end{equation} where $N_{L}$ is the number of grains in the layer. The evolution of this mean vertical displacement $\bar{dy}$ just after the first tap is presented in Figure \ref{im:deltaYAfterTap1}. During and after the tap, we do observe neither upward movement nor take off from the bottom of the container. The movement of the grains situated on the top of the pile is well fitted by a free fall $\bar{d y} = -\rm{g}t^2/2$, with $\rm{g} = 9.81$ $m/s^2$, for $0 \; ms < t < 10$ ms. After $t = 10$ ms, the movement is damped. This damping is fitted by an exponential decay. The damping is due to friction and collisions with both other grains in the packing and the side walls. For grains situated in the center and in the bottom of the pile, the amplitude of the mean vertical displacement is small. Moreover, the dynamics is slower. Indeed, due to the Janssen effect, the influence of the friction with the side walls is more important. The fact that the five curves in Figure \ref{im:deltaYAfterTap1} do not coincide for large time is due to a densification witch depends on the depth in the packing. It should be noted that this tap is the first one of the series. For the followings taps, the displacements of the grains becomes much smaller and similar curves cannot be measured when the packing fraction increases.

\section{General observations}

At the macroscopic scale, compaction is measured through the packing fraction $\rho$. The pile is made of $N$ cylindrical grains of length $\ell$ and of diameter $d$. Therefore, the packing fraction is the ratio between the total surface of the grains $N \ell d$ on the picture and the total surface of the pile $S_{pile}$. The later is estimated by taking into account the average position of the surface grains. One has 

\begin{equation}
\rho = \frac{N \ell d}{S_{pile}}.
\end{equation}

\begin{figure}[h]
\begin{center}
\includegraphics[scale=0.5]{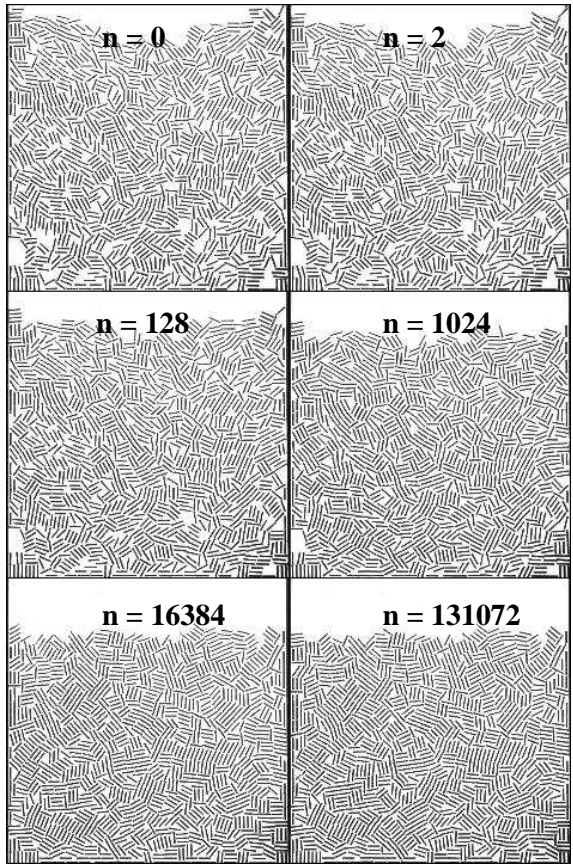} 
\end{center}
\caption{Evolution of the pile during a compaction experiment. Different stages of the compaction are illustrated. For $n=0$, an angular disorder is observed. When $n$ increases, some domains of aligned grains appear, and they grow thereafter.}
\label{im:montageCylExp}
\end{figure}

Figure \ref{im:montageCylExp} presents snapshots of the pile during compaction. The packing fraction of the initial pile ($\rho^{Cyl}_0 = 0.775$) is small in comparison with the same experiment made with spherical particles ($\rho^{Sph}_0 = 0.825$) \cite{LumayVandewalle2}. Furthermore, an angular disorder is clearly observed for $n=0$. When $n$ increases, some domains of parallel grains appear and grow. The degree of disorder decreases. This growth of ordered domains could be compared to a crystallization process. The final packing fraction, at the end of the experimental run, is $\rho^{Cyl}_{\infty} = 0.877$. For spherical particles in the same conditions, the final packing fraction is close to $\rho^{Sph}_{\infty} = 0.862$ \cite{LumayVandewalle2}. The range $\rho_{\infty} - \rho_0$ of the packing fraction accessible to anisotropic particles is thus quite large with respect to spherical particles.

We do not observe a preferential orientation of the grains along the vertical sides or along the horizontal bottom wall. Indeed, the aspect ratio of the grains $\alpha = 3.1$ is too small to observe a preferred horizontal orientation as reported in \cite{prolatePacking}. Moreover, the ratio between the pile width and the grain length $L/\ell = 22$ is too large to observe a preferred vertical orientation as reported in \cite{rods}. In our experimental configuration, we observe only local angular organization due to neighbouring grains oriented along their long axis.


\section{Compaction curves}

\begin{figure}[htbp]
\begin{center}
\includegraphics[scale=0.6]{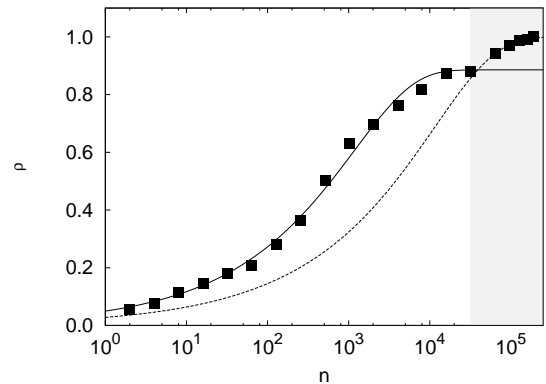} 
\end{center}
\caption{A typical compaction curve giving the normalized packing fraction $\tilde{\rho}$ as a function of the tap number $n$. The first ($n < 3 . 10^4$) and the second ($n > 3 . 10^4$) stages of the compaction process are fitted using our crystallization model \cite{LumayVandewalle2}, i.e. Eq. (\ref{eq:AvramiRhoLaw}). Error bars correspond to the size of the symbol used in this plot. The second stage of the compaction process is colored in grey.}
\label{im:densiteCylExp}.
\end{figure}

The evolution of the normalized packing fraction $\tilde{\rho}$ as a function of the tap number $n$ is presented in Figure \ref{im:densiteCylExp}. The normalized evolution is quite reproducible. We observe a change of the compaction dynamics in the experimental compaction curve above $\tilde{\rho} \approx 0.8$, i.e. after $n \approx 10^4$ taps. Such a jump in the packing fraction evolution $\rho(n)$ has been reported in earlier 3D experiments \cite{rods} as quoted in the introduction. This effect was not observed in the case of sphere packings. Two compaction processes seems to occur at different stages. The first stage and the second stage of the compaction curve are well fitted by two independent ``Avrami laws'' (\ref{eq:AvramiRhoLaw}) proposed in our earlier work \cite{LumayVandewalle2} with the exponents $\gamma_{1} = 0.75 \pm 0.05$ and characteristic times $\tau_{1} = 2200 \pm 200$, $\tau_{2} = 20000 \pm 2000$. $\gamma_{2} $ is has been fixed to 0.75. The characteristic time of the second stage of the compaction process is ten times larger than the characteristic time of the first stage. Therefore, the dynamics of the second stage is very slow in comparison with the first stage dynamics. Furthermore, the Avrami exponents are superior to 0.5. This can be explained by the existence of a favoured direction for the displacement of the grains. The presence of two stages can be understood as follows. In the first stage, the compaction is essentially due to fast spatial reorganisations (translations) of the grains. The second stage corresponds to slow orientational reorganisations (rotations) of the grains. This interpretation could be guessed in the snapshots of Figure \ref{im:montageCylExp} and this will be confirmed by both microscopic and mesoscopic studies in the next sections.

\section{Grain organization}

The aim of this section is to study the formation and the growth of mesoscopic structures, i.e. domains made of grains ideally packed. We attempt to relate this growth mechanism to the compaction dynamics.

\begin{figure}[htbp]
\begin{center}
\includegraphics[scale=0.2]{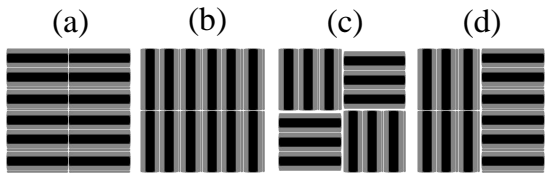} 
\end{center}
\caption{Four different possible arrangements of grains ($\alpha = 3$) leading to the ideal packing fraction $\rho = 1$.}
\label{im:idealDomains}
\end{figure}

For spherical particles in a 2D system \cite{LumayVandewalle2}, the only one ideal packing is the hexagonal compact arrangement. The corresponding packing fraction for 2D pile is $\rho \approx 0.91$. For cylindrical particles, the ideal packing could be reached using various structures. In Figure \ref{im:idealDomains}, many different organizations of anisotropic ($\alpha = 3$) grains lead to the unique ideal packing fraction $\rho = 1$. Beside two trivial configurations (Fig \ref{im:idealDomains}a and Fig \ref{im:idealDomains}b) in which grains are perfectly aligned, the ideal packing fraction is reached when domains of aligned grains are formed and well arranged. Therefore, anisotropic particles have to search for one among many ideal arrangement in a more complex way than spherical particles. In this respect, the grain anisotropy $\alpha$ is an important parameter when looking at compaction. However, a study of compaction for different $\alpha$ values is outside the scope of this paper.

\begin{figure}[htbp]
\begin{center}
\includegraphics[scale=0.4]{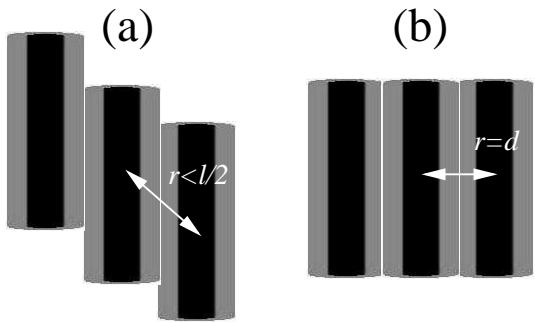} 
\end{center}
\caption{The sketch (a) presents a domain of three aligned grains. The sketch (b) presents a domain of three ideally ordered grains.}
\label{im:grainsAR-R}
\end{figure}

Let us study the grain organization in the packing. We define two fractions of ordered grains : the fraction of aligned grains $\phi_{a}$ (Figure \ref{im:grainsAR-R}a) and the fraction of ideally ordered grains $\phi_{io}$ (Figure \ref{im:grainsAR-R}b). The second type of grain ordering forms blocks. In order to measure these fractions, we use a simple criterion. For each grain in the pile, we search for its nearest neighbours. If at least one neighbour has the same orientation and is distant of a half grain length $\ell/2$ (respectively distant of a grain diameter $d$) from the grain, it is considered to belong to an aligned domain (respectively ideally ordered domain). Finally, we obtain the number $N_{a}$ of grains aligned (respectively the number $N_{io}$ of grains ideally ordered). The fraction of grains aligned $\phi_{a}$ (respectively the fraction of grains ideally ordered $\phi_{io}$) is the ratio between the number of grains aligned $N_{a}$ (the number of grains ideally ordered $N_{io}$) and the total number  $N$ of grains in the pile. One has $\phi_{a} = N_{a}/N$ and $\phi_{io} = N_{io}/N$.

\begin{figure}[htbp]
\begin{center}
\includegraphics[scale=0.6]{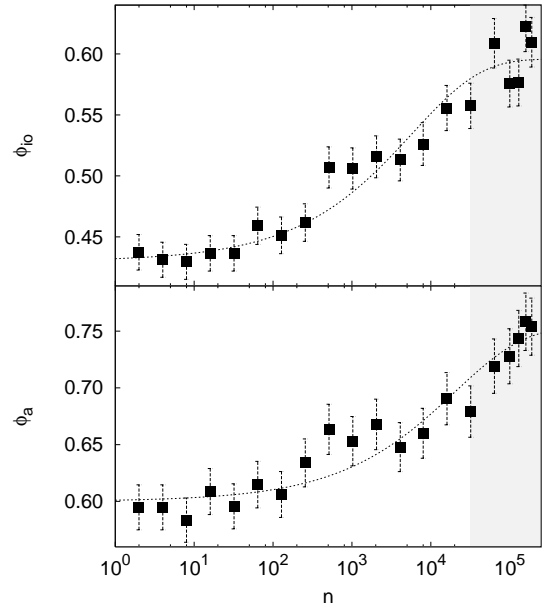} 
\end{center}
\caption{Evolution of both fractions of grains aligned $\phi_{a}$ and fraction of grains ideally ordered $\phi_{io}$ as a function of the tap number $n$. The curves are fits using Eq. (\ref{eq:AvramiModelAlign}) for $\phi_{a}$ and Eq. (\ref{eq:AvramiModelOrder}) for $\phi_{io}$. The grey area of the figure correspond to the second stage observed in the compaction curve (Figure \ref{im:densiteCylExp})}
\label{im:phiCylExp}
\end{figure}

Figure \ref{im:phiCylExp} presents the fraction of aligned grains $\phi_{a}$ and the fraction of grains ideally ordered $\phi_{io}$ as a function of the number of taps $n$. The error bars correspond to the standard deviation observed when we repeat the experiment many times. The fluctuations of both fractions $\phi_{a}$ and $\phi_{io}$ do not come from measurement imprecisions, but from natural fluctuations in the system. Both fractions increase with $n$, meaning that domains form and grow during compaction. Since the ideal ordering is more restrictive than grain alignment, $\phi_{io}$ is always below $\phi_{a}$. Both experimental data are well fitted by the Avrami law \cite{avrami}. One has
\begin{equation}
\label{eq:AvramiModelAlign}
\frac{\phi_{a} - \phi_{a,0}}{\phi_{a,\infty} - \phi_{a,0}} = 1-\exp \left[ - \left( \frac{n}{\tau_{a}}\right) ^ {\gamma} \right],
\end{equation} and
\begin{equation}
\label{eq:AvramiModelOrder}
\frac{\phi_{io} - \phi_{io,0}}{\phi_{io,\infty} - \phi_{io,0}} = 1-\exp \left[ - \left( \frac{n}{\tau_{io}}\right) ^ {\gamma} \right],
\end{equation} The experimental data is fitted by these equations with an exponent $\gamma$ fixed to 0.75 (i.e. the value obtained with the fit of the compaction curve) and with characteristic times $\tau_{a} = 20315 \pm 5000$ and $\tau_{io} = 5811 \pm 1500$. The initial fractions are $\phi_{a,0} = 0.6$ and $\phi_{io,0} = 0.43$. The final fractions are given by the fit $\phi_{a,\infty} = 0.72 \pm 0.05$ and $\phi_{io,\infty} = 0.60 \pm 0.05$ respectively. It is remarkable that the long time $tau_{a}$ is similar to the long relaxation time obtained in density curves. 

\begin{figure}[htbp]
\begin{center}
\includegraphics[scale=0.6]{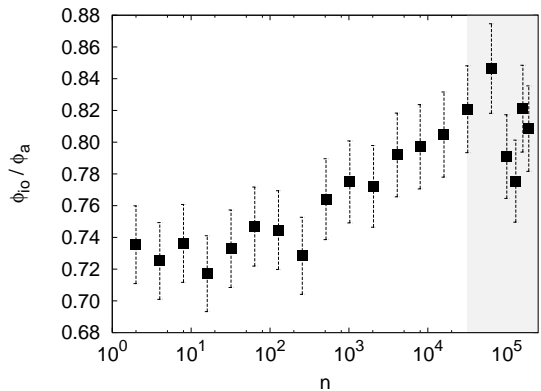} 
\end{center}
\caption{Ratio between the fraction of grains ideally ordered $\phi_{io}$ and the fraction of grains aligned $\phi_{a}$ as a function of the tap number $n$. The grey area of the figure correspond to the second stage observed in the compaction curve (Figure \ref{im:densiteCylExp})}
\label{im:RapportPhiCylExp}
\end{figure}

The ratio between both fractions $\phi_{io}/\phi_{a}$ (Figure \ref{im:RapportPhiCylExp}) allow us to measure the proportion of grains ideally ordered. During the first stage of compaction (for $n < 3 \; 10^4$), this ratio increases with the tap number $n$. This increase could be interpreted as follows. At the beginning of the process, the domains are formed of aligned grains (Figure \ref{im:grainsAR-R}a). Afterwards, the domains are sheared to form domains ideally ordered (Figure \ref{im:grainsAR-R}b). During the second stage of compaction (for $n > 3 \; 10^4$), the ratio $\phi_{io}/\phi_{a}$ start to fluctuate. Indeed, both fractions $\phi_{io}$ and $\phi_{a}$ are close to the saturation.

\begin{figure}[htbp]
\begin{center}
\includegraphics[scale=0.6]{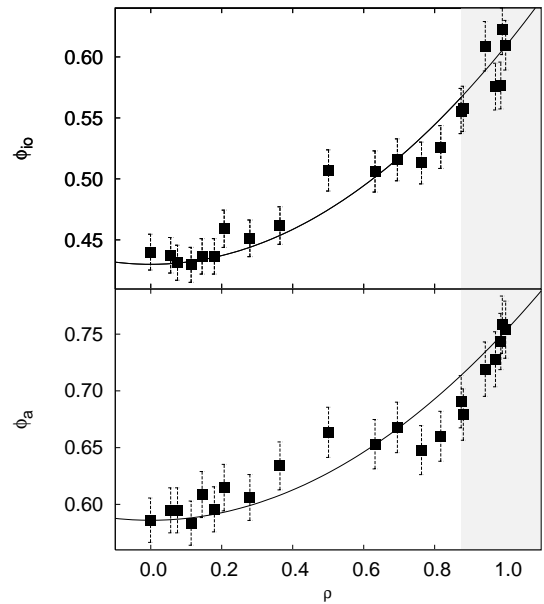} 
\end{center}
\caption{The fraction of grains aligned $\phi_{a}$ (respectively, fraction of ideally ordered grains  $\phi_{io}$) as a function of the normalized packing fraction $\tilde{\rho}$. The data is fitted by a parabolic law. The grey area of the figure correspond to the second stage observed in the compaction curve (Figure \ref{im:densiteCylExp})}
\label{im:phiVSrhoCylExp}
\end{figure}

Figure \ref{im:phiVSrhoCylExp} presents both fractions of grains aligned $\phi_{a}$ and fraction of grains ideally ordered $\phi_{io}$ as a function of the normalized packing fraction $\tilde{\rho}$. Of course, both fractions increase with the packing fraction. The growth seems to be parabolic as in our previous work \cite{LumayVandewalle2}. Indeed, in two dimensions, the growth of the packing fraction is related to the slow diffusion of domain boundaries. Data on Figure \ref{im:phiVSrhoCylExp} is fitted with such a parabolic law.

\section{Grain mobilities}

In this section, local displacements are studied during the compaction process. Since all grain positions are recorded, it is possible to measure the vertical displacement $\delta y_i$, the horizontal displacement $\delta x_i$ and the angle variation  $\delta \theta_i$ of each grain $i$ after a single tap or after a series of taps. A dimensionless average translational mobility $\mu_t$ can be defined from the ratio between particle displacements $\sqrt{(\delta x_i)^2 + (\delta y_i)^2}$  and the length of the cylinders $\ell$. One has
\begin{equation}
\mu_t = \frac{1}{N} \sum_{i=1}^{N} {\sqrt{(\delta x_i)^2 + (\delta y_i)^2} \over \ell}.
\end{equation}

During the compaction process, grains also experience rotations. A dimensionless average rotational mobility $\mu_t$ can also be defined from the ratio between particle extremity displacements $\ell \theta_i$ and the length of the cylinders $\ell$. One has

\begin{equation}
\mu_r = \frac{1}{N} \sum_{i=1}^{N} {(\delta \theta_i)}.
\end{equation}

\begin{figure}[htbp]
\begin{center}
\includegraphics[scale=0.6]{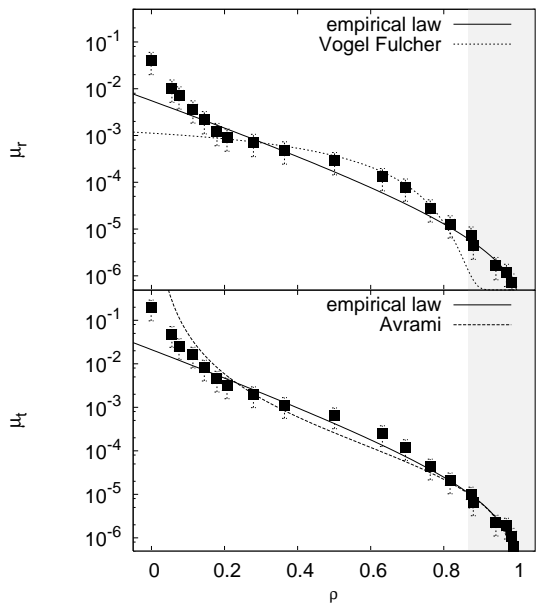} 
\end{center}
\caption{The translational and rotational mobilities $\mu_t$ and $\mu_r$ as a function of the normalized packing fraction $\tilde{\rho}$. The continuous curves are fits using the empirical law (Eq \ref{eq:mobilite},\ref{eq:mobiliteRot}). The dashed curves are fits using the Vogel-Fulcher law for the rotational mobility and using the Avrami law for the translational mobility. The grey area of the figure correspond to the second stage observed in the compaction curve (Figure \ref{im:densiteCylExp})}
\label{im:mobiliteCylExp}
\end{figure}

Figure \ref{im:mobiliteCylExp} presents the decrease of the translational and the rotational mobilities $\mu_t$ and $\mu_r$ as a function of the normalized packing fraction $\tilde{\rho}$. The shape of the two curves is similar. In order to avoid a surcharge of the figure we do not fit both mobilities by the laws Eq. (\ref{eq:vogelFulcher}), (\ref{eq:AvramiMuLaw}) and (\ref{eq:EmpiricalMuLaw}). Both mobilities are fitted by the empirical law. One has
\begin{equation}
\label{eq:mobilite}
\mu_{t} = \mu^{0}_{t} (1-\tilde{\rho}) e^{-a_{t} \tilde{\rho}},
\end{equation} and 
\begin{equation}
\label{eq:mobiliteRot}
\mu_{r} = \mu^{0}_{r} (1-\tilde{\rho}) e^{-a_{r} \tilde{\rho}},
\end{equation} where $\mu^{0}_{t} = 0.022 \pm 0.004$ and $\mu^{0}_{r} = 0.010 \pm 0.002$ are the initial translational and rotational mobilities respectively. Caging parameters $a_{t} = 6.64 \pm 0.38$ and $a_{r} = 5.36 \pm 0.37$ are the decaying rates of both mobilities. The decrease of the rotational mobility is also fitted by the Voguel-Fulcher law Eq. (\ref{eq:vogelFulcher}) and the translational mobility is also fitted by the the Avrami law Eq. (\ref{eq:AvramiMuLaw}). The shape of the curve is quite well fitted by the Avrami law. 

\begin{figure}[htbp]
\begin{center}
\includegraphics[scale=0.6]{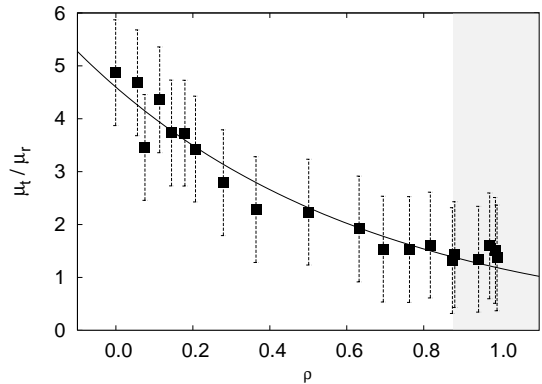} 
\end{center}
\caption{Ratio between the translational mobility $\mu_t$ and the rotational mobility $\mu_r$ as a function of the normalized packing fraction $\tilde{\rho}$. The curve is a fit by an exponential decay $ \exp(-\Delta a \tilde \rho)$ with $\Delta a = 1.4 \pm 0.1$. The grey area of the figure correspond to the second stage observed in the compaction curve (Figure \ref{im:densiteCylExp})}
\label{im:RapportMobiliteCylExp}
\end{figure}

The ratio between the translational and the rotational mobilities $\mu_t / \mu_r$ as a function the normalized packing fraction $\tilde{\rho}$ is presented in figure \ref{im:RapportMobiliteCylExp}. For low densities ($\tilde{\rho} < 0.4$), the compaction is mainly related to translational moves. For high packing fraction, the proportion of rotational moves increases. This transition (translational to rotational moves) could explain the change of dynamics compaction observed in the compaction cureve (Figure \ref{im:densiteCylExp}).

\section{Conclusion}

In summary, we have measured three physical quantities during the granular compaction of cylindrical particles. They correspond to three different scales in the system. At the macroscopic scale, we have observed two stages during the compaction process. Both stages could be fitted by the ``Avrami law'' \cite{LumayVandewalle2}. The presence of two dynamical stages could be explained by the mesoscopic and by the microscopic analysis. At the mesoscopic scale, we have observed the formation and the grow of domains made of aligned grains. Our experiments suggest that after formation, such domains are sheared to order ideally. At the microscopic point of view, the measurements of both translational and rotational mobilities reveal that the beginning of the compaction process is essentially related to translational motion of the grains. At the late stages of compaction, the grain rotations are more pronounced and drive the process. Therefore, the first stage corresponds to a fast spatial reorganisation of the grains to forms domains of aligned grains. The second stage corresponds to a slow orientational reorganisation of the grains to form domains of ideally ordered grains.


\section{Acknowledgements}

This work has been supported by the contract ARC 02/07-293. The authors thank C. Becco, S. Dorbolo, H. Caps, E. Cl\'ement, F. Ludewig an P. Richard for valuable discussions.


\begin{thebibliography}{2} 
\bibitem{deGenne} P. G. de Gennes, Rev. Mod. Phys. \textbf{71}, S374 (1999)
\bibitem{granularSolidLiquidsGases} H. M. Jaeger, S. R. Nagel and R. P. Behringer, Rev Mod Phys, \textbf{68}, 1259 (1996)
\bibitem{Duran} J. Duran, \textit{Sables, poudres et grains} (Eyrolles Sciences, Paris, 1999)
\bibitem{Kudrolli} A. Kudrolli, Rep. Prog. Phys. \textbf{67}, 209 (2004)
\bibitem{Francois} F. Ludewig and N. Vandewalle, EPJE \textbf{18}, 367 (2005)
\bibitem{Glasses} A. Coniglio, A. Fierro, H. Herrmann and M. Nicodemi, \textit{Unifying concepts in granular media and glasses} (Elsevier,2004)
\bibitem{vanMegen} P.N. Pusey and W. van Megen, Nature \textbf{320}, 340 (1986)
\bibitem{densFluc} E. R. Nowak, J. B. Knight, E. Ben-Naim, H. M. Jaeger, and S. R. Nagel, Phys. Rev. E \textbf{57}, 1971 (1998)
\bibitem{densityRelax} J. B. Knight, C. G. Fandrich, Chun Ning Lau, H. M. Jaeger, and S. R. Nagel, Phys. Rev. E \textbf{51}, 3957 (1995)
\bibitem{richard} P. Richard, M. Nicodemi, R. Delannay, P. Ribi\`ere, and D. Bideau, Nature Materials \textbf{4}, 121 (2005)
\bibitem{kww1} P. Philippe, and D. Bidau, Europhys. Lett. \textbf{60}, 677 (2002) 
\bibitem{LumayVandewalle2} G. Lumay and N. Vandewalle, Phys. Rev. Lett. \textbf{95}, 028002 (2005)
\bibitem{Tetris} E. Caglioti, V. Loreto, H.J. Herrmann and M. Nicodemi, Phys. Rev. Lett. \textbf{79}, 1575 (1997)
\bibitem{Tetris2} F. Ludewig, S. Dorbolo and N. Vandewalle, Phys. Rev. E \textbf{70}, 051304 (2004)
\bibitem{slowRelax} E. Ben-Naim, J. B. Knight, E. R. Nowak, H. M. Jaeger, and S. R. Nagel, Physica D \textbf{123}, 380 (1998)
\bibitem{boutreux} T. Boutreux and P.G. de Gennes, Physica A \textbf{244}, 59 (1997) 
\bibitem{arenzon1} Y. Levin, J.J. Arenzon and M. Sellitto, EuroPhys. Lett. \textbf{55}, 767 (2001)
\bibitem{arenzon2} J.J. Arenzon, Y. Levin and M. Sellitto, Physica A \textbf{325}, 371 (2003)
\bibitem{clusterModel} K. L. Gavrilov, Phys. Rev. E \textbf{58}, 2107 (1998).
\bibitem{Kudrolli2} D.L. Blair, T. Neicu, and A. Kudrolli, Phys. Rev. E \textbf{67}, 031303 (2003)
\bibitem{PG2005Lum} G. Lumay and N. Vandewalle, Powders \& Grains 2005
\bibitem{Tshape} F. Ludewig, N. Vandewalle and S. Dorbolo, Gran. Mat. \textbf{8}, In press (2006)
\bibitem{LumayVandewalle1} G. Lumay and N. Vandewalle, Phys. Rev. E \textbf{70}, 051314 (2004)
\bibitem{convectAnisotropic} P. Ribi\`ere, P. Richard, R. Delannay and D. Bideau, Phys. Rev. E \textbf{71}, 011304 (2005)
\bibitem{rods} F. X. Villarruel, B. E. Lauderdale, D. M. Mueth, and H. M. Jaeger, Phys. Rev. E \textbf{61}, 6914 (2000)
\bibitem{prolatePacking} K. Stokely, A. Diacou, and Scott V. Franklin, Phys. Rev. E \textbf{67}, 051302 (2003)
\bibitem{thin-rod} A.P. Philipse and A. Verberkmoes, Physica A \textbf{235}, 186 (1997)
\bibitem{spherocylinders} S.R. Williams and A.P. Philipse, Phys. Rev. E \textbf{67}, 051301 (2003)
\bibitem{pierrePhilippe} P. Philippe, \textit{Etude th\'eorique et exp\'erimentale de la densification des milieux granulaires}, PhD thesis, Universit\'e de Rennes 1, (2002)
\bibitem{avrami} M. Avrami, J. Chem. Phys. \textbf{7}, 1103-1112 (1939); M. Avrami, J. Chem. Phys. \textbf{8}, 212 (1940).

\end{thebibliography}
\end{document}